\documentclass[preprint,eqsecnum,aps]{revtex4}

\usepackage{graphicx}
\usepackage{dcolumn}
\usepackage{amsmath}

\makeatletter

\begin{document}

\title{Studies of parton thermalization at RHIC}

\author{Ghi R. Shin}
\affiliation{Department of Physics, Andong National University,
                    Andong, South Korea}
\author{Berndt M\"uller}
\affiliation{Department of Physics, Duke University,
                    Durham, NC 27708-0305, USA}

\date{\today}

\begin{abstract}
We consider the evolution of a parton system which is formed in the 
central region just after a relativistic heavy ion collision. The 
parton consist of mostly gluons, minijets, which are produced by 
elastic scattering between constituent partons of the colliding nuclei.
We assume the system can be described by a semi-classical Boltzmann 
transport equation, which we solve by means of the test particle 
Monte-Carlo method including retardation. The partons proliferate
via secondary radiative $gg \rightarrow ggg$ processes until the 
thermalization is reached for some assumptions.
The extended system is thermalized at about $t=1.6$ fm/$c$ with
$T = 570$ MeV and stays in equilibrium for about $2$ fm/$c$ with 
breaking temperature $T = 360$ MeV in the rapidity central region.
\end{abstract}

\maketitle

\section{Introduction}

It is generally expected that a Quark-Gluon Plasma(QGP) can be formed
after a relativistic heavy ion collision even at RHIC energy. However, 
it has not been clear yet how a QGP is formed in a heavy ion collision,
as well as what the equation of state is, if it is formed. Some of 
these questions, which are important for the identification of signals 
of the QGP and of the properties of dense strongly interacting matter, 
are the subjects of our study.

Beginning with the moment of contact of two colliding heavy ions, 
initially coherent quanta, which can be identified as partons, from 
the projectile nucleus collide with those from the target nucleus and
can become incoherent particle excitations. They can, as well, produce 
additional particles by radiative processes \cite{bjo83,bla87}.
There may be several different mechanisms responsible for particle 
production in relativistic heavy ion collision depending on the 
transverse momentum of the produced particles. Those of high transverse 
momentum, which is greater than a threshold momentum $p_0$, are usually 
called minijets and can be approximately calculated by the methods 
of perturbative QCD (pQCD) using the parton structure functions of the 
colliding nuclei. On the other hand, the ``soft'' partons of low 
transverse momentum, with $p_T \leq p_0$, are thought to be produced by 
the semi-classical decoherence of strong, random gluon fields contained
in the colliding nuclei, sometimes called color glass condensate 
\cite{mcl94,am00,knv00}.

While the primary partons of high transverse momentum (minijets) are 
produced near the central rapidity region and have relatively high 
transverse energy, the soft partons are distributed over a wide range 
of rapidities and have lower transverse energy.  These soft partons 
can play a role as a thermal bath or background field, which slows down
the minijets by inducing them to transfer a substantial fraction of their 
energy to the bath, thus enhancing their chance to be thermalized. It was
pointed out\cite{bir93,bai00}, however, that the major mechanism of energy
deposition is through radiative ($2 \rightarrow 3$) processes, suggesting
that it is sufficient to consider a minijet system in order to explore 
the chemical and thermal equilibration of the system. This is what we
are doing here.

We study the initial phase-space distribution of minijets assuming no 
correlation between momentum and space and another one with correlation, 
especially those given by Cooper at al\cite{coo02}, which we will call 
CMN distribution, in Section II. We briefly describe the semi-classical 
Boltzmann equations of motion and the Monte-Carlo simulations to solve 
the equations in Section III. We present the results of our numerical 
solutions and a summary in Section IV.

\section{Initial Distributions}

The hard gluons produced after a relativistic heavy ion collision have 
been studied extensively by many authors. Its momentum distribution as 
well as the total number of partons can be calculated by the method of 
pQCD. To see the dominant effect at very early times after the onset of 
a relativistic heavy ion collision, we consider only minijet gluons.
Using the parton distribution of colliding nucleus $A$, 
$f_{i/A}(x,Q^2) = f_{i/N}(x,Q^2) R_A(x,Q^2)$, where $f_{i/N}(x,Q^2)$ is 
the parton distribution of a free nucleon and $R_A(x,Q^2)$ the nuclear 
ratio function, the minijet distribution can be calculated by the 
$2 \rightarrow 2$ minijet cross section per nucleon in a $A+A$ collision,
\begin{eqnarray}
{dN^{jet}}\over{dp_T dy} &=& K T(b) \int dy_2 {{2\pi p_T} \over{\hat s}}
\sum_{ij,\,kl} x_1 f_{i/A}(x_1,p_T^2) x_2 f_{j/A}(x_2,p_T^2)
\sigma_{ij\rightarrow kl}(\hat{s},\hat{t},\hat{u}),
\label{pt_y_dis}
\end{eqnarray}
where $K$ denotes a factor to include the higher order diagrams, which 
we will set to $K=2$ throughout our study. $T(b)$ is the nuclei geometric 
factor at an impact parameter $b$. We use the GRV98 set of parton 
distributions for a free nucleon \cite{grv98} and the EKS98 parametrization 
for the ratio function \cite{eks99}. $x_1$ and $x_2$ are the Bjorken 
scaling variables of parton $i$ and $j$ in nucleons of two colliding nuclei, 
respectively. $y_1$ and $y_2$ are the rapidities of scattered or produced
partons. 

The $gg$ scattering cross section at leading order is
\begin{eqnarray}
\sigma_{gg\rightarrow gg} &=& {{9 \pi
\alpha_s^2}\over{2\hat{s}}}[3-{{\hat{t}\hat{u}}\over{\hat{s}^2}}
-{{\hat{u}\hat{s}}\over{\hat{t}^2}} -
{{\hat{s}\hat{t}}\over{\hat{u}^2}} ],
\end{eqnarray}
where the ``hat'' symbol on the Mandelstam variables indicates those 
of a parton.  The relations between the variables are given by
\begin{eqnarray}
x_1 &=& p_T (e^{y_1} + e^{y_2}) /\sqrt{s},\\
x_2 &=& p_T (e^{-y_1} + e^{-y_2}) /\sqrt{s},\\
\hat{s} &=& x_1 x_2 s, \\
\hat{t} &=& -p_T^2 (1 +e^{y_2 - y_1}),\\
\hat{u} &=& -p_T^2 (1 +e^{y_1 - y_2}).
\end{eqnarray}
We also rewrite the available kinematic region for convenience \cite{ham99},
\begin{eqnarray}
p_0\, ^2 \leq p_T\, ^2 \leq ({{\sqrt{s}}\over{2 \cosh y}})^2, \\
-\log ({{\sqrt{s}}\over{p_T}} - e^{-y}) \leq y_2 \leq 
\log({{\sqrt{s}}\over{p_T}} - e^{-y}),\\
|y| \leq \log ( {{\sqrt{s}}\over{2 p_0}} + \sqrt{
{{s}\over{4p_0\,^2}} - 1}).
\end{eqnarray}

We consider only a head-on collision ($b=0$) and use the leading-order 
expression for the running coupling constant 
$\alpha_s = 4 \pi [b_0\log (Q / \Lambda_{\rm QCD})]^{-1}$, 
where $Q$ is the momentum transfer, $b_0 = 11-{2\over 3}n_f$, and 
$\Lambda_{\rm QCD} = 200$ MeV and $n_f = 3$.  The production of minijet
gluons strongly depends on the transverse momentum cut $p_0$. We use 
$p_0 = 1$ GeV at RHIC energy, which is marginal for the use of pQCD. 
Note that we set the minimum transverse momentum 
to be equal to the saturation momentum, where the gluon distribution
becomes semi-classical \cite{knv00}. The total numbers of minijets 
(integrated over rapidity) is about 4800 at RHIC.
These numbers are quite large since we set 
the momentum cutoff for the minijets at the lower limit of the range
$p_0^2 \sim 1 - 2$ GeV$^2$ at RHIC.  The rapidity distribution is 
shown in Fig. 1 of reference \cite{coo02}.

To obtain the phase space distribution of minijets, we first assume 
that there are no correlation between the momentum and space coordinates 
of a produced parton. The momentum distribution can be given by the 
minijet distribution, Eq. (\ref{pt_y_dis}),
\begin{eqnarray}
f(p_T, y) &=& C {1 \over p_T^2} {{dN^{jet}}\over{dy \,dp_T}}.
\end{eqnarray}
We can choose $(p_T,y)$ using this distribution and the azimuthal 
angle $\phi$ with equal weight between $(0,2\pi)$ by the Monte-Carlo 
method, by which we select the energy and momentum of the produced parton.
The spatial distribution of minijets can be deduced from the fact that 
the minijets are produced from elastic scattering between constitutient 
partons of two colliding nuclei. Assuming the space-time point of produced 
particle is just that of elastic scattering, we can estimate the space-time 
positions based on the classical picture as follows. We assume that the 
probability for having an elastic scattering of constituent partons is 
proportional to the nuclear density overlap of the colliding nuclei. 
The transverse position of elastic scatterings is thus obtained by
\begin{eqnarray}
x &=& r \sin \theta \cos \phi, \\
y &=& r \sin \theta \sin \phi,
\end{eqnarray}
where the distribution of constant nuclear density in the nuclear rest 
frame is $P(r,\theta,\phi) = 3r^2 / (4\pi R^3)$.

To obtain the longitudinal position and the time of a collision, we 
consider a sphere of radius $R$ which has a constant density and 
is moving with a constant velocity $v\approx c$.  We assume two 
identical spheres suffering a head-on collision in the CM frame. 
At a given transverse collision position $(x,y)$, we consider 
longitudinal tubes of area $\Delta A$ through the point $(x,y)$. 
Because of the Lorentz contraction, the half-length of the tube
is $D/\gamma$ with $D=\sqrt{R^2 -x^2 -y^2}$ and $\gamma=(1-v^2/c^2)^{-1/2}$.
The tubes from both nuclei begin to overlap one another starting 
at $t_s = (R-D)/\gamma v$ and ending at $t_e = (R+D)/\gamma v$.
Note that we set the time of first contact between the two colliding 
nuclei at $t=0$.  At time $t$, the overlap volume is given by
\begin{eqnarray}
\Delta V &=& 2 \Delta A\, v (t-t_s)/\gamma.
\label{overlap_v}
\end{eqnarray}
The maximun overlap volume is $\Delta V = 2 \Delta A\,D/\gamma$ 
at $t=R/\gamma v$. The probability density which elastic parton 
collisions occur is proportional to the overlap volume and the product
of the two Lorentz contracted parton densities:
\begin{eqnarray}
P(t) &=& {{\gamma^2v^2}\over{2D^2}}(t-t_s)
\end{eqnarray}
Using this collision time distribution, we select the collision time 
randomly:
\begin{eqnarray}
t&=& t_s + {{2D}\over{\gamma v}} \sqrt{a_1},
\label{rand_t}
\end{eqnarray}
where $a_1\in [0,1]$ is a random number. Once we have $t$, we 
can calculate the overlap volume, Eq.(\ref{overlap_v}), and its longitudinal 
range, $- v(t-t_s) < z < +v(t-t_s)$.  We again choose the longitudinal 
location $z$ of the collision in this range randomly:
\begin{eqnarray}
z&=& v [2 (t-t_s) a_2 - (t-t_s)] ,
\end{eqnarray}
where $a_2\in [0,1]$ is another random number. 

We also consider the phase-space distributions (\ref{pt_y_dis}) at RHIC 
which were recently proposed by Cooper, Mottola and Nayak\cite{coo02} 
(CMN Distribution). 
The boost noninvariant distribution, for example, is given
\begin{eqnarray}
f(p_T,\eta, y,\tau_0) &=& C {dN^{\rm jet}\over{dydp_T}}
{{e^{-{{(y-\eta)^2}/{\sigma^2(p_T)}}}\over
{\tau_0 p_T^2\sigma(p_T) e^{\sigma^2(p_T)/4}}}},
\end{eqnarray}
where $\eta$ is spatial rapidity. $\sigma(p_T)$ is an unknown parameter 
function, which relates the space-rapidity correlation to the transverse 
momentum and was chosen to be constant ($\sigma^2 = 0.28$) at RHIC energy 
in Ref. \cite{coo02}.  Using this distribution function, we can randomly 
selects the phase-space coordinates $(p_T,y,\eta)$ of each produced parton. 
We assume in this distribution that the parton production time is given by 
Eq.~(\ref{rand_t}).

\begin{figure}
\includegraphics{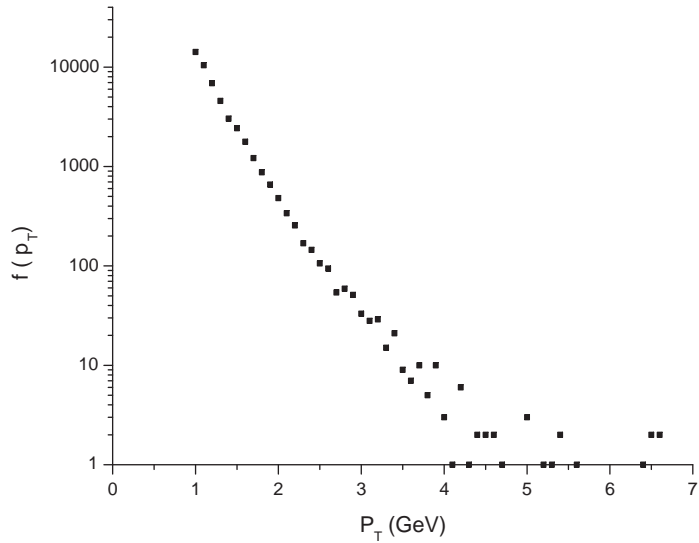}
\caption{The transverse momentum distribution of a Monte Carlo sampled 
parton system as a function of transverse momentum in GeV unit.} 
\label{fig1}
\end{figure}

Figure \ref{fig1} shows the transverse momentum distribution of Monte-Carlo 
sampled partons with a minimal transverse momentum $p_0 = 1$ GeV/$c$.
Most of the produced particles have the transverse monenta less than 
3 GeV/$c$ and the number of produced gluons sharply increases as the 
transverse momentum gets smaller.

\begin{figure}
\includegraphics{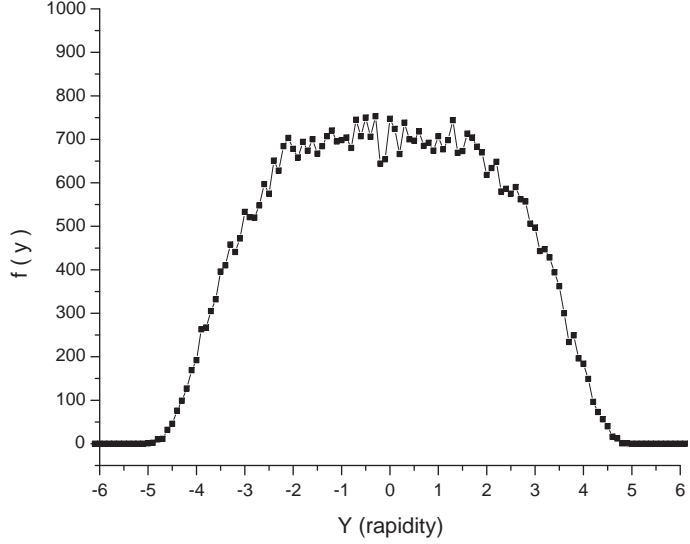}
\caption{The rapidity distribution of the produced partons.} 
\label{fig2}
\end{figure}

Figure \ref{fig2} shows the rapidity distribution of Monte-Carlo 
sampled particles, which can be compared to the rapidity distribution 
given in Ref. \cite{coo02}. The distribution shows that the rapidity 
is flat in the central region, but falls off for $|y| > 3$.
\\
Since we use Monte-Carlo sampling of the distribution, we use several 
different realizations of the initial distribution (events) and average 
over the events to get the physical values.

\section{Numerical Solution}

The initial-state gluons, which we will call primary gluons from now 
on, evolve according to the Boltzmann 
equation\cite{shin02, gro80,cse94,gei92,gei95,zpc},
\begin{widetext}
\begin{eqnarray}
p^\mu \partial_\mu f_{g} (x, \vec p ) &=&  \int_2\int_3\int_4
\frac{1}{2}W_{gg \rightarrow gg} [ f_g(3)f_g(4) - f_g(1)f_g(2) ] \nonumber\\ 
&+& \int_2\int_3\int_4 W_{gq \rightarrow g q} [ f_g(3)f_q(4)
- f_g(1)f_q(2)] \nonumber \\
&+& \int_2\int_3\int_4\int_5 \frac{1}{6}W_{gg \rightarrow ggg} 
[f_g(4)f_g(5)- f_g(1)f_g(2) ] \nonumber \\
&+& \int_2\int_3\int_4 W_{g \bar q
\rightarrow g \bar q}[ f_g(3)f_{\bar q}(4) - f_g(1)f_{\bar
q}(2) ] \nonumber\\
&+& \int_2\int_3\int_4 \frac{1}{2}[ W_{q \bar q \rightarrow g g}
f_q(3)f_{\bar q}(4) - W_{g g\rightarrow q \bar q }f_g(1)f_g(2)],
\label{BTE1}
\end{eqnarray}
\end{widetext}
where we used the abbreviated notations $f_g(i) = f_g(\vec x, \vec
p_i; t)$, $q =(u, d, s)$, $\bar q = (\bar u, \bar d, \bar s)$, and
$\int_i \equiv \int d {\vec p_i}/E_i$. The transition rates can 
be expressed as
\begin{eqnarray}
W_{gg_1 \rightarrow g'g_1'} = s \sigma(s,\Theta)
\delta^{(4)}(p+p_1-p'-p_1') .
\end{eqnarray}
Note that we assign the label $1$ to the momentum $\vec p$.
We explicitly include the particle symmetry factor in the
classical limit. The $ggg \rightarrow gg$ process is not included
in the equation of motion since the system of high energy partons expands
very rapidly so that the available volume increases and the channel
is suppressed strongly.

On the other hand, the secondary quarks and antiquarks produced in 
parton collisions satisfy the equations of motion,
\begin{widetext}
\begin{eqnarray}
p^\mu \partial_\mu f_q (x, \vec p ) &=& \int_2\int_3\int_4 C\,
W_{qq' \rightarrow q q'} [ f_q(3)f_{q'}(4) - f_q(1)f_{q'}(2) ]\nonumber \\
&+& \int_2\int_3\int_4 W_{gq \rightarrow g q} [ f_q(3)f_g(4) -
f_q(1)f_g(2) ] \nonumber \\
&+& \int_2\int_3\int_4 \frac{1}{2} [ W_{gg \rightarrow q\bar q}\,
f_g(3)f_g(4) - W_{q\bar q \rightarrow gg}\, f_q(1)f_{\bar q}(2) ].
\label{BTE2}
\end{eqnarray}
\begin{eqnarray}
p^\mu \partial_\mu f_{\bar q}(x, \vec p ) &=& \int_2\int_3\int_4
C\, W_{\bar q \bar q' \rightarrow \bar q \bar q'} [ f_{\bar
q}(3)f_{\bar q'}(4) - f_{\bar q}(1)f_{\bar q'}(2) ] \nonumber \\
&+& \int_2\int_3\int_4 W_{g \bar q \rightarrow g \bar q} [ f_{\bar
q}(3)f_g(4) - f_{\bar q}(1)f_g(2) ] \nonumber \\
&+& \int_2\int_3\int_4 \frac{1}{2} [ W_{gg \rightarrow q\bar q}\,
f_g(3)f_g(4) - W_{q\bar q \rightarrow gg}\, f_{\bar q}(1)f_q(2) ].
\label{BTE3}
\end{eqnarray}
\end{widetext}
where $C = 1/2$, if the final state consists of identical particles
and $C = 1$ otherwise. We neglect the quantum mechanical effects from 
Bose enhancement factors $(1+f_g)$ for a gluon in the final state and 
Pauli blocking factors $(1-f_q)$ for final-state quarks or antiquarks.

We use the parton cascade code, which was developed by us and described
in ref.~\cite{shin02}, to solve the equations of motion for a given initial
state. We first initialize all primary partons in space-time with their
four-momenta and look for the next possible collision.
If a collision occurs, we select the collision products and their new 
momenta according to the total and differential cross section, respectively, 
at the space-time of relativistic maximum force point between the colliding 
particles. This procedure is terminated when the evolution time of the
cascade is greater than the final time set as an input parameter.

\section{Results and Conclusion}

%
%

\begin{figure}
\includegraphics{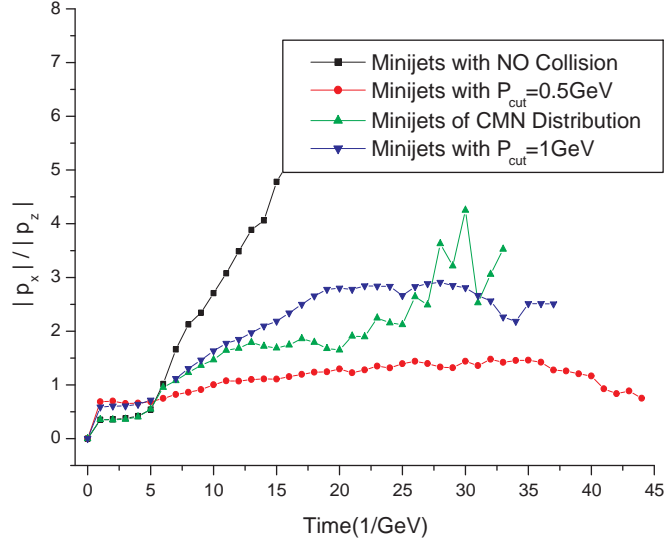}
\caption{The ratio ${{\sum |p_x|} \over{ \sum |p_z }}$ 
in a small test sphere ($r=1.1$ fm$^3$) at center region.} 
\label{fig4}
\end{figure}

Fig. \ref{fig4} shows the ratio $\sum |p_x| / \sum |p_z|$ in a small 
sphere with radius $R=1.1$ fm at the center of the coordinate frame 
as a function of time, where 
\begin{eqnarray}
\sum | p_x | &=& \int d^3x \int {d^3p} {{|p_x|}\over{E}} f(x,p).
\end{eqnarray}
When this ratio becomes unity, an isotropic, equilibrated configuration 
has been reached.

As expected, the ratio for the evolution without collisions remains clearly 
different from unity, which means that there is no kinetic equilibration. 
The ratio with collisions shows the kinetic isotropy from a time of about 
8 GeV$^{-1} = 1.6$ fm/$c$  to 17 GeV$^{-1} = 3.4$ fm/$c$ for the system 
initialized with minijet partons. The simulation includes small angle 
elastic scattering, for which we set the minimum momentum transfer to 
$p_{\rm cut} = 0.5$ GeV/$c$. We also show the results of simulations with 
hard scattering only, $p_{\rm cut} = 1$ GeV/$c$. Even though the total 
number of partons in the system is almost same with 9700, only the system 
with small angle scatterings reaches a kinetically isotropic state. 

On the other hand, the ratio for the system of minijets of CMN distribution 
always remains away from unity. We can understand this result as follows: 
Since the value of the spatial rapidity of a produced parton is close to 
its momentum rapidity, those partons which have positive longitudinal 
momentum, will be produced with positive longitudinal position, the right 
hand side of the collision plane, and the parton with negative longitudinal 
momentum will be born at a negative longitudinal position, on the left-hand 
side of the collision plane. This correlation thus makes possible only 
those collisions between the particles which are running in the same 
longitudinal direction but different transverse momentum. This phenomena 
reduces the number of collisions and the collision energy substantially, 
so that the evolution is essentially free streaming.

\begin{figure}
\includegraphics{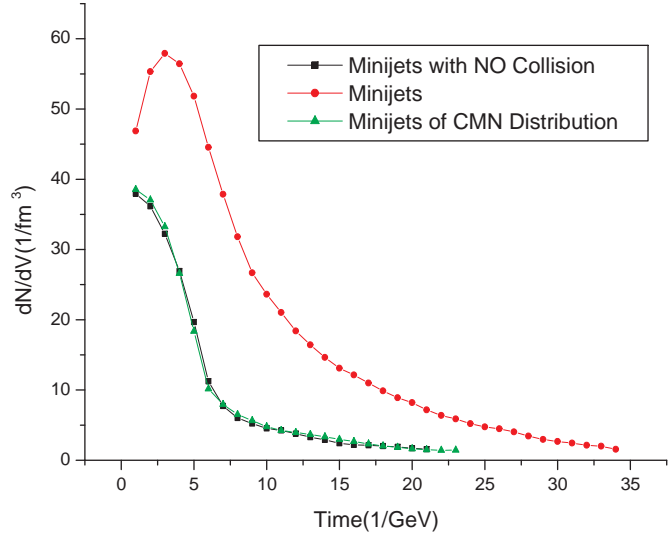}
\caption{The particle density $N/V$ in a small test sphere at central region.} 
\label{fig5}
\end{figure}

Figure \ref{fig5} shows the number density in the sphere as function of time.
The density of partons for minijets simulations decreases faster than $1/t$ 
and from 32 fm$^{-3}$ at $t=8$ GeV$^{-1}$ to 11 fm$^{-3}$, which is much 
higher than the density of valence quarks of a free nucleon, at $t=17$ 
GeV$^{-1}$ in the small sphere at the center. The density equals to the 
valence quark density of a free nucleon at $t = 34$ GeV$^{-1}$. Note that 
we plot the data if the particle density of the sphere is greater than that 
of valence quarks in a nucleon, 1.4 fm$^{-3}$.

\begin{figure}
\includegraphics{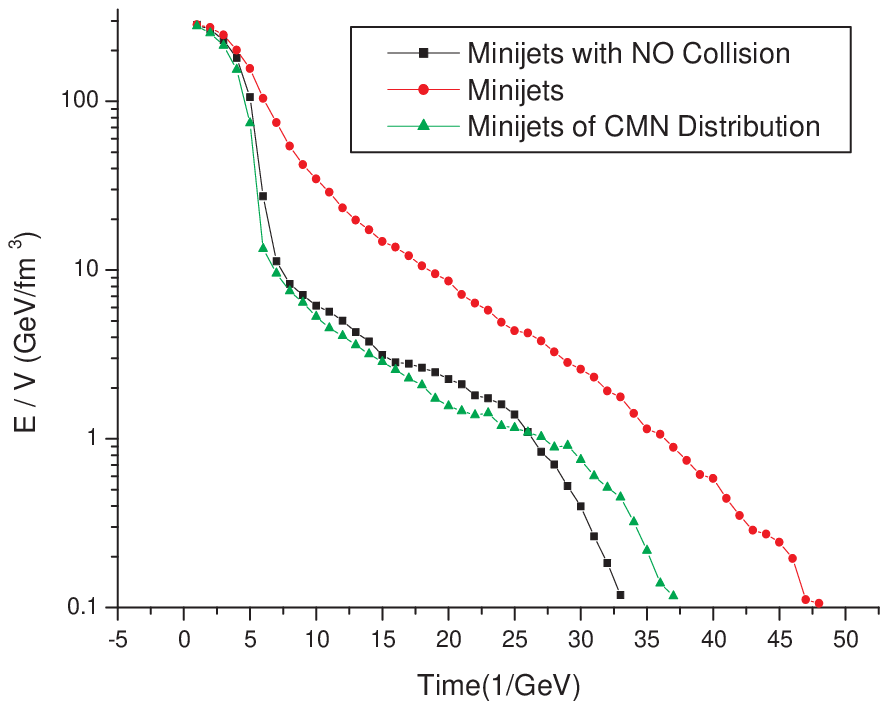}
\caption{The energy density $E/V$ of the sphere.} 
\label{fig6}
\end{figure}

Figure \ref{fig6} shows the energy density of the sphere as a function of time,
which is almost exponential decreasement. The energy density decreases from 
54 GeV/fm$^3$ at $t=8$ GeV$^{-1}$ to 12 GeV/fm$^3$ at $t=17$ GeV$^{-1}$ for 
our minijet simulations.

\begin{figure}
\includegraphics{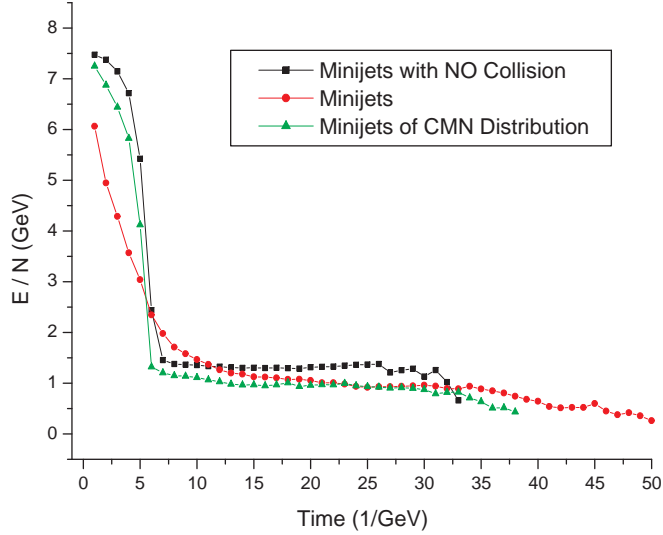}
\caption{The energy per particle $E/N$ in the sphere.} 
\label{fig7}
\end{figure}

Figure \ref{fig7} shows the energy per parton in the sphere as a function 
of time.  The energy per parton of free streaming(no collision) is much 
higher at early stage of expansion than those of interacting systems, 
which means the parton loses a substantial amount of energy via scattering.
The energy decreases from 1.7 GeV at $t=8$ GeV$^{-1}$ to 1.1 GeV at 
$t=17$ GeV$^{-1}$ for a minijet system. If we assume from Fig \ref{fig4} 
that the local system of the sphere at the center is thermalized, the 
relativistic gas has the temperature $T = 566$ MeV at $t=8$ GeV$^{-1}$ 
to 367 MeV at $t=17$ GeV$^{-1}$ for a minijet system and a mixed system, 
respectively. This system is much hotter than the QGP transition 
temperature $T_c\approx 170$ MeV.

\begin{figure}
\includegraphics{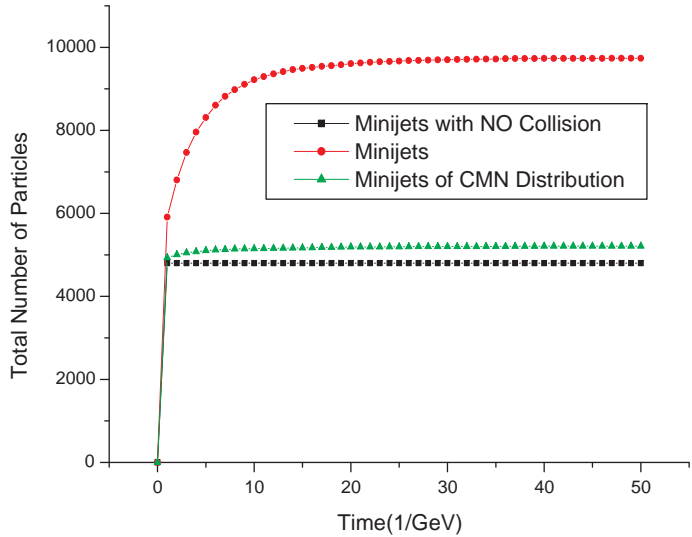}
\caption{The total number of particles in the system as function of time.} 
\label{fig8}
\end{figure}

Figure \ref{fig8} shows the total number of particles in the system. The 
secondary particles are produced within 2 fm/$c$ after the first contact of 
two colliding nuclei. This shows that the chemical freeze-out comes much 
earlier than the thermal freeze-out even at the center of collision. The 
total number of particles is about 9700 for a minijet system and 5200 for
a minijet system of CMN distribution.  As many authors have pointed out, 
the $gg \rightarrow ggg$ process, even though the process has the kinematical 
constraint, $\theta(p_T \lambda_f - cosh y)\theta(\sqrt{s}/2 - p_T cosh y)$,
which is $E_{\rm cm} > 1.54$ GeV in our simulations, plays a dominant role in 
depositing energy and thermalization of a system.

\begin{figure}
\includegraphics{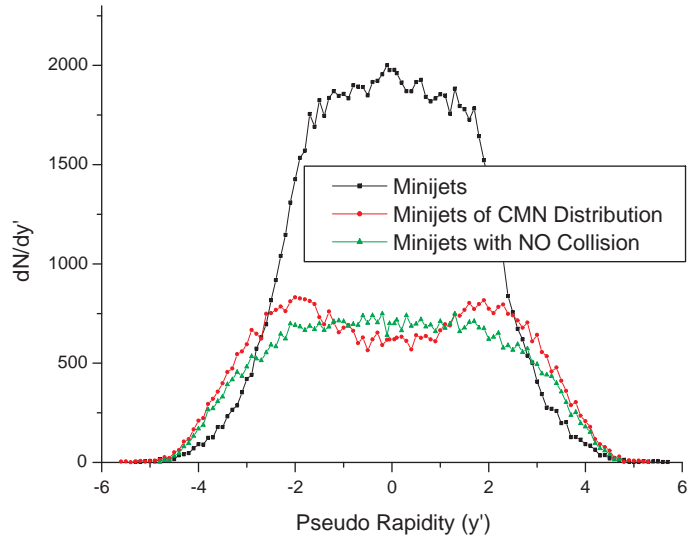}
\caption{The pseudo rapidity distribution of parton systems as functions
of time.} 
\label{fig9}
\end{figure}

Fig. \ref{fig9} shows the pseudo rapidity distribution of all of partons in 
the system at $t=50$ GeV$^{-1}$.  Since partonic collisions are ceased by 
this time, the distribution will be preserved to the final state unless the 
hadronic interaction changes them. We can see that the rapidity distribution 
is narrowed since the high rapidity partons at early time, Fig.~\ref{fig2},
have been scattered to produce the low rapidity partons. We also point out 
that the rapidity distribution is different from the final state (hadronic) 
rapidity distribution in number and the width\cite{jac02}.

We conclude that the system is thermalized even with minijets at RHIC energy 
if the small angle scatterings are included.  When we look at the small 
sphere of radius $R=1.1$ fm at center, the system is thermalized from 
$t=1.6$ fm/$c$ till about 3.4 fm/$c$.  And the temperature is decreasing 
from 566 to 367 MeV and the particle density from 32 to 11 fm$^{-3}$.
Even after this time, the system is still hot untill about 8 fm/$c$ at 
the center and may cool down further after hadronization.  The evolution 
of the system, which has no correlation between position and momentum, 
is very similar in character to the "bottom-up thermalization" scenario
of Baier {\it et al.}\cite{bai00}, in which the hard gluons liberated from 
the colliding nuclei within $\tau \approx Q_s^{-1}$ collide among 
themselves to produce soft gluons and thermalize via small angle scattering.

{\it{$^*$Acknowledgements}}: G. R. Shin was supported in part by grant 
No. 2000-1-111000-007-1 from the Basic Research Program of the Korea 
Science and Engineering Agency.

\end{document}